\documentclass[12pt]{article}
\pdfoutput=1
\usepackage[utf8]{inputenc}
\usepackage{amsmath}
\usepackage{amssymb}
\usepackage{units}
\usepackage{graphicx}
\usepackage{slashed}
\usepackage{array}
\usepackage[leftcaption,raggedright]{sidecap}
\usepackage{caption}
\usepackage[hidelinks]{hyperref}

\newcommand\pubnumber{NuPhys2023-Aleena-Rafique}
\newcommand\pubdate{\today}

\def\napoli{Argonne National Laboratory}

\def\Title#1{\begin{center} {\Large #1 } \end{center}}
\def\Author#1{\begin{center}{ \sc #1} \end{center}}
\def\Address#1{\begin{center}{ \it #1} \end{center}}

\newcommand\pubblock{\rightline{\begin{tabular}{l} \pubnumber\\
         \pubdate  \end{tabular}}}
\newenvironment{Abstract}{\begin{quotation}  }{\end{quotation}}
\newenvironment{Presented}{\begin{quotation} \begin{center} 
             PRESENTED AT\end{center}\bigskip 
      \begin{center}\begin{large}}{\end{large}\end{center} \end{quotation}}

\def\beq{\begin{equation}}
\def\eeq#1{\label{#1}\end{equation}}
\def\eeqn{\end{equation}}

\def\beqa{\begin{eqnarray}}
\def\eeqa#1{\label{#1}\end{eqnarray}}
\def\eeqan{\end{eqnarray}}

\let\bar=\overbar

\def\Dslash{\not{\hbox{\kern-4pt $D$}}}
\def\dslash{\not{\hbox{\kern-2pt $\del$}}}

\def\msb{{\bar{\ssstyle M \kern -1pt S}}}

\begin{document}
\begin{titlepage}
\pubblock

\vfill
\Title{Neutrino energy scale measurements for final state interactions using advanced computing in DUNE }
\vfill
\Author{ Aleena Rafique}
\Address{\napoli}
\Author{ for the DUNE collaboration}
\vfill
\begin{Abstract}
The Deep Underground neutrino experiment (DUNE)~\cite{dune}, consisting of near (DUNE-ND)~\cite{dune-nd} and far (DUNE-FD)~\cite{dune-fd} detectors, is a long-baseline experiment that is designed to measure neutrino oscillations, as well as searches beyond the standard model. The DUNE-FD will operate with an active volume of 40 kiloton liquid argon and will be situated at Sanford Underground Research Facility (SURF) in South Dakota. The DUNE-ND will be placed close to the neutrino source and measure an un-oscillated neutrino beam for precise measurement of oscillation parameters. Final State Interactions (FSI) are the secondary interactions of the daughter particles of the neutrino with other nucleons within the same argon nucleus. We present the impact of using different final state interaction models on the neutrino energy scale measurements in DUNE using advanced computing at Argonne National Laboratory.
\end{Abstract}
\vfill
\begin{Presented}
NuPhys2023, Prospects in Neutrino Physics\\
King's College, London, UK,\\ December 18--20, 2023
\end{Presented}
\vfill
\end{titlepage}
\def\thefootnote{\fnsymbol{footnote}}
\setcounter{footnote}{0}

\section{Introduction}
\subsection{DUNE}
DUNE is a next-generation, long-baseline (1300 km) neutrino oscillation experiment which will carry out a detailed study of neutrino mixing utilizing high-intensity $\nu_\mu$ and $\bar{\nu_\mu}$ beams measured over a long baseline. DUNE consists of an FD having 40 kiloton active mass liquid argon time projection chamber (LArTPC) at Sanford Underground Research Facility (SURF) located in South Dakota, and an ND that will be situated at Fermilab. DUNE will measure neutrino oscillation probability to determine mass ordering and charge-parity violation phase via $\bar{\nu_e}$ appearance and $\bar{\nu_\mu}$ disappearance. DUNE will also carry out a rich program of the search for BSM physics and supernova neutrinos. 

The studies presented in this document focus on DUNE-FD. The DUNE-FD consists of four LArTPC modules each having a fiducial mass of 10 kiloton placed at SURF. One of these module is a horizontal-drift LArTPC; the other will be a vertical-drift TPC. The technology R$\&$D for the 3rd and 4th module technology is ongoing.

\begin{figure}[h!]
\centering
\includegraphics[width=0.5\linewidth]{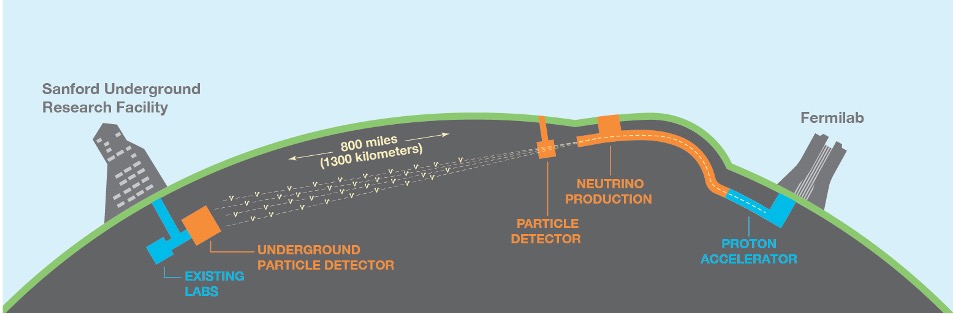}
\includegraphics[width=0.3\linewidth]{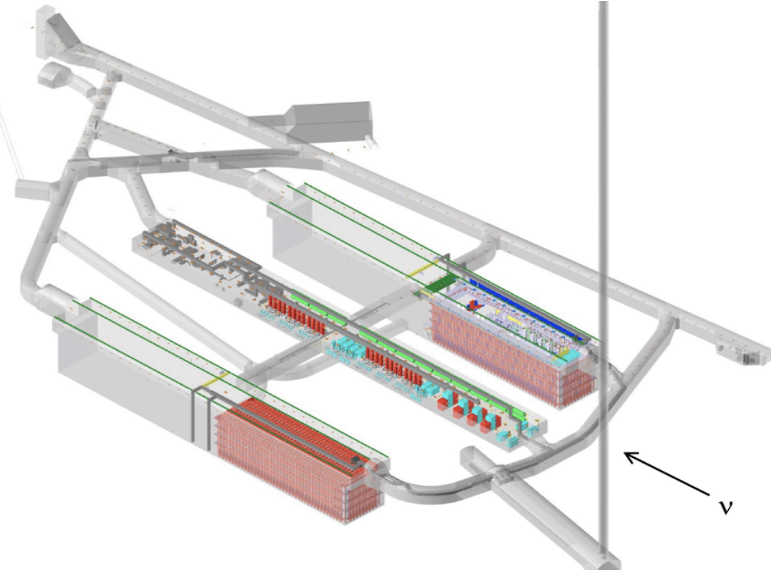}
\caption{The DUNE experiment schematic showing the near and far detectors at Fermilab and SURF respectively~\cite{dune-fd} (left); the DUNE FD site~\cite{dune-fd} (right).}
\label{fig:dune}
\end{figure}

Figure~\ref{fig:dune}~(left) presents the DUNE experiment schematic; and Figure~\ref{fig:dune}~(right) presents the DUNE-FD site at SURF with a cavern for four TPCs. 

\subsection{Argonne Computing Resources}
\begin{figure}[h!]
\centering
\includegraphics[width=0.5\linewidth]{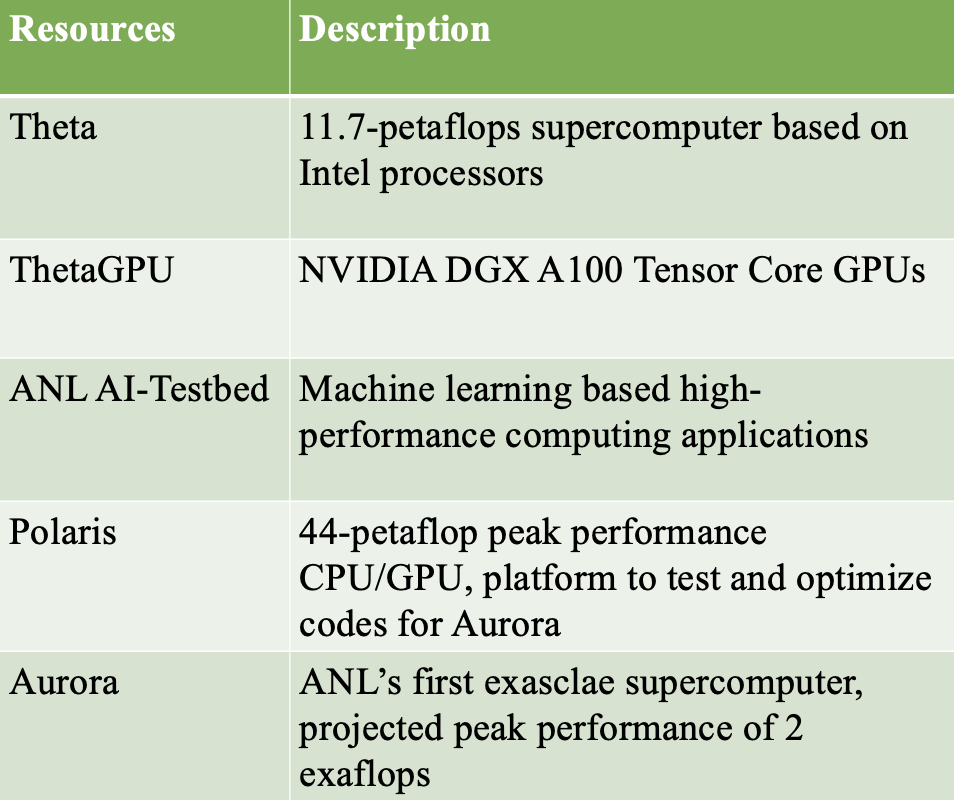}
\includegraphics[width=0.4\linewidth]{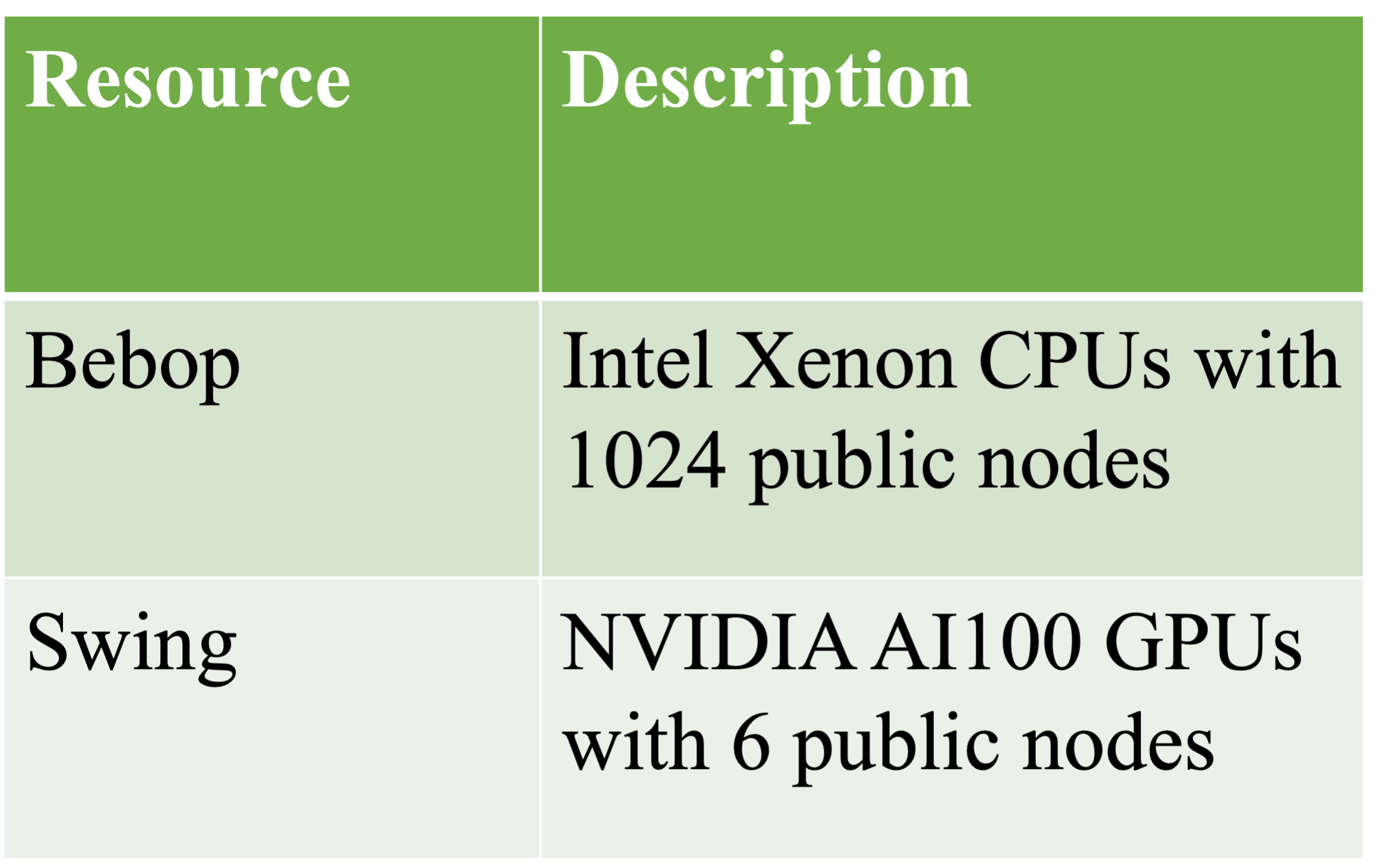}
\caption{Description of the resources present at ALCF (left); at LCRC (right).}
\label{fig:computing_chart}
\end{figure}

The generation and management of the scientific datasets are central to achieving the scientific objectives of DUNE. The DUNE detectors are expected to collect several terabytes of data every second, a volume that presents significant computational challenges. Our work on the Argonne Supercomputing resources focuses on producing the neutrino detector simulation data for physics analysis.  In this paper we present results on the effects of final state interactions in the initial event generation stage, which generates several gigabytes of data and consumes hundreds of CPU and GPU combined hours per production.  We will also extending this effort to include detector simulation which is substantially more CPU intensive. The goals include scaling down the simulation time and adapting the data storage methods for efficient I/O on HPC systems. Argonne has two computing facilities, namely Argonne Leadership Computing Facility (ALCF) and Laboratory Computing Resource Center (LCRC). Figure~\ref{fig:computing_chart} presents a list and description of all the available computing resources present at these facilities. We have been using bebop 
 and swing machines for the work presented here in this document. 

\subsection{Final State Interactions} 
\begin{figure}[h!]
\centering
\includegraphics[width=0.3\linewidth]{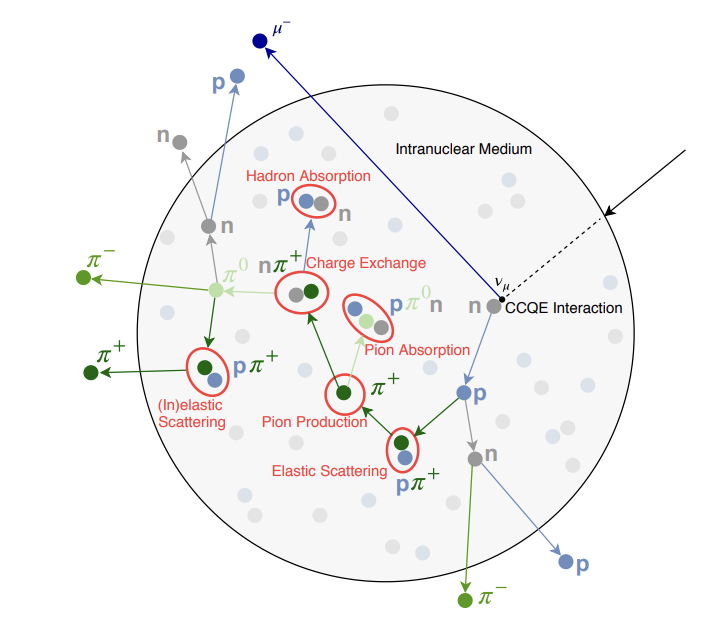}
\includegraphics[width=0.6\linewidth]{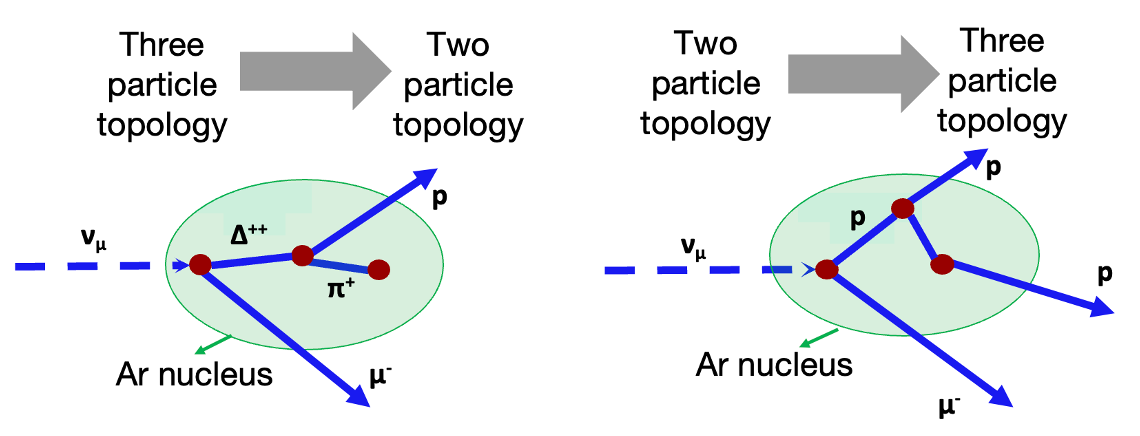}
\caption{Figure from~\cite{fsi} illustrating the FSIs (left); FSI changing the topologies of the interaction (right).}
\label{fig:FSI}
\end{figure}

When a neutrino interacts with an argon nucleus, the initial state particles are produced. The initial state hadrons undergo secondary interactions, called the final state interactions (FSI), with other nucleons present within the same nucleus.  Figure~\ref{fig:FSI}~(left) presents an illustration of FSIs, in which a proton (light-blue line) undergo various types of hadronic FSIs~\cite{fsi}. 

FSI present an important way to mask the identity of the primary interaction and can change the topology of the interaction and can also impact the final state energy. FSI are dominant in heavier nuclei such as argon. Figure~\ref{fig:FSI}~(right) presents how initially a three-particle topology neutrino interaction can be changed into a two-particle topology and vice versa. 

\section{Sample Generation and Workflow}
5k events were generated using GENIE (version 3.4 AR23\_20i)~\cite{GENIE} standalone neutrino event generator using ANL computing resources. Each same initial state interaction was then propagated to the following FSI models. 

\begin{itemize}
\item $hA$: the default model used in most current neutrino simulations. It only
considers one hadron rescattering.
\item $hN$: it considers multiple rescatterings until the hadron escape the nucleus.
\item $INCL$++: the entire hadron-residual system changes through time steps.
\item $Geant4$: Bertini Cascades (G4BC)~\cite{G4BC}, more sophisticated model.
\end{itemize}

As a result of this, four samples were generated. The impact on the final state energies is presented in this document.  

\section{Observations and Results}
\begin{figure}[h!]
\centering
\includegraphics[width=0.4\linewidth]{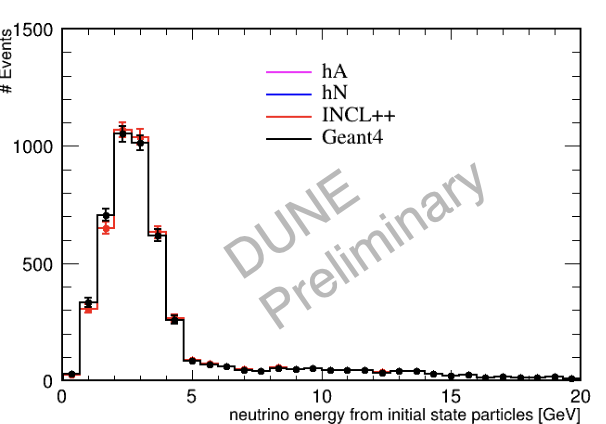}
\includegraphics[width=0.4\linewidth]{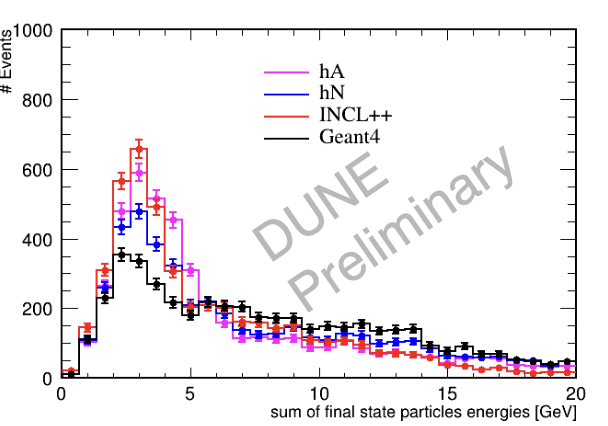}
\caption{Energy sum of the initial state particles (left); Energy sum of the final state particles (right).}
\label{fig:energies}
\end{figure}

The sum of all the initial and final state energies is calculated by
\begin{equation}
E_{i/f} = E_h + E_l - E_n
\end{equation}

where $E_{i}$ or $E_{f}$ is the sum of the initial state (before FSI) or final state (after FSI) particle energies respectively; $E_h$ is the initial or final state hadronic energy sum; $E_l$ is the primary lepton energy; and $E_n$ is the hit nucleon energy. The sum is over all the particles of an interaction. This energy spectrum is presented in Figure~\ref{fig:energies} (left) for initial state and (right) for final state energy sum for all FSIs. We see that the initial state energies are consistent across all four FSIs as expected. We also see that there is a discrepancy in final state energy sum from the default tune (hA) as large as ~45\%. These discrepancies limit our model understanding and will impact the energy scale and reconstruction. 

\begin{figure}[h!]
\centering
\includegraphics[width=0.4\linewidth]{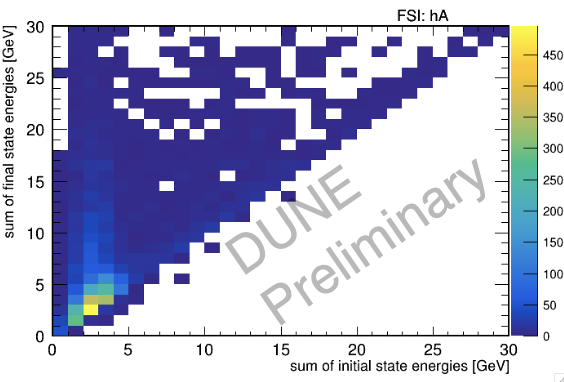}
\includegraphics[width=0.4\linewidth]{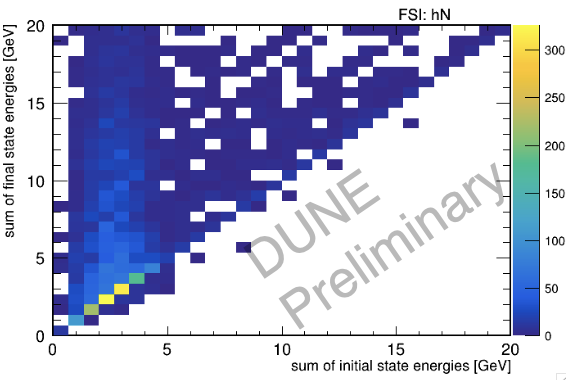}
\includegraphics[width=0.4\linewidth]{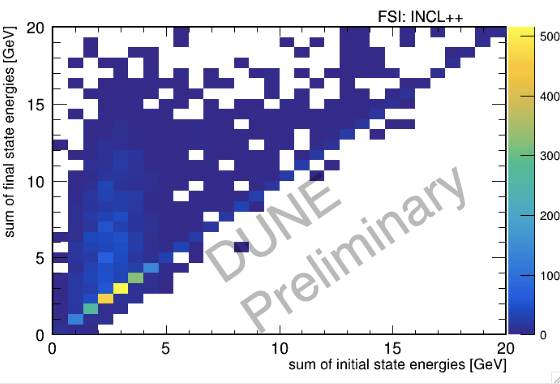}
\includegraphics[width=0.4\linewidth]{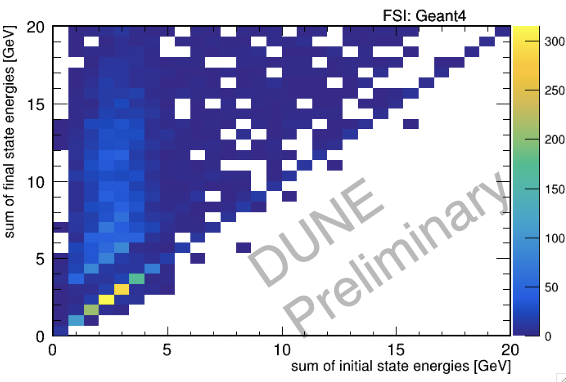}
\caption{Initial versus final state energy sum for $hA$ (top left); $hN$ (top right); $INCL$++ (bottom left); and $Geant4$ (bottom right).}
\label{fig:energies_2D}
\end{figure}

Figure~\ref{fig:energies_2D} presents the initial versus final state energies for different FSI models. There seems to be a better agreement in initial and final state energies for the models $hN$ and $INCL$++ compared to the default $hA$ model.

\section{Conclusions and Outlook}
This is the first demonstration of utilizing Argonne computing for DUNE physics studies. We observed how FSI can impact the neutrino energy spectrum. In future, we plan to look into the dependence of the energy difference between different neutrino interaction types (QE, RES, DIS etc). We also plan to reconstruct the neutrino energy using FD reconstruction tools. In addition, we will calculate the effect of these uncertainties on the CP violation sensitivity studies.

\providecommand{\href}[2]{#2}\begingroup\raggedright\endgroup

\end{document}